\documentstyle[11pt]{article}

 \newfont{\goth}{cmbx9}
 \newfont{\gothi}{cmbx9}

 \newcommand{\smin}{\,\raisebox{0.06em}{${\scriptstyle \in}$}\,}
 \newcommand{\smsubset}{\,\raisebox{0.06em}{${\scriptstyle \subset}$}\,}

 \newcommand{\gotg}{\mbox{\goth g}}
 \newcommand{\gotig}{\mbox{\gothi g}}

 \newcommand{\gotigh}{\mbox{\gothi h}}

 \newcommand{\smotimes}{\,{\scriptstyle \otimes}\,}

\begin{document}
\title{New classical r-matrices from integrable non-linear sigma models\\~
       \\
       Contribution presented at the XIX ICGTMP Salamanca}

\author{\underline{~~J. Laartz$\,^1$ ~,~~}\\
        M. Bordemann$\,^2$~,~~
        M. Forger$\,$~,~~
        U. Sch\"aper$\,$}
 \footnotetext[1]{Supported by Studienstiftung des Deutschen Volkes}
 \footnotetext[2]{Supported by Deutsche Forschungsgemeinschaft, Contract
                  No.~Ro 864/1-1}

 \date{\normalsize
       Fakult\"at f\"ur Physik der Universit\"at Freiburg, \\
       Hermann-Herder-Str.~3, D-7800 Freiburg / FRG }

\maketitle
\thispagestyle{empty}
\begin{abstract}
\noindent
Non linear sigma models on Riemannian symmetric spaces constitute the most
general class of classical non linear sigma models which are known to be
integrable.
Using the the current algebra structure of these models
their canonical structure is analysed and it is shown that their non ultralocal
fundamental Poisson bracket relation is governed by a field dependent non
antisymmetric r-matrix obeying a dynamical Yang Baxter equation. \newline
The fundamental Poisson bracket relations and the r-matrix are derived
explicitly
and a new kind of algebra is found that is supposed to replace the
classical Yang Baxter algebra governing the canonical structure of ultralocal
models.\newline
\end{abstract}

\vfill
 \begin{flushright}
 \parbox{12em}
 { \begin{center}
 University of Freiburg \\
 THEP 92/20 \\
 September 1992
 \end{center} }
 \end{flushright}

\newpage
\renewcommand{\thefootnote}{\arabic{footnote}}
\setcounter{page}{1}

\noindent {\Large\bf{New classical r-matrices from integrable non-linear
           $\sigma$-models}}  \newline
\begin{center}
{\large \bf {\underline{J. Laartz}}, M.Bordemann, M.Forger, U.Sch\"aper\\
 Universit\"at Freiburg, Fakult\"at f\"ur Physik, Hermann-Herder Str.3,
 D-7800 Freiburg}
 \end{center}

\noindent
{\small {\bf Abstract:}
Non-linear sigma models on Riemannian symmetric spaces constitute the most
general class of classical non-linear sigma models which are known to be
integrable. Using the the current algebra structure of these models
their canonical structure is analyzed and it is shown that their non-ultralocal
fundamental Poisson bracket relation is governed by a field dependent non
antisymmetric r-matrix obeying a dynamical Yang Baxter equation.
The fundamental Poisson bracket relations and the r-matrix are derived
explicitly
and a new kind of algebra is found that is supposed to replace the
classical Yang Baxter algebra governing the canonical structure of ultralocal
models.}
\vspace{0.7cm}

\noindent To analyze the (classical) structure of two-dimensional integrable
field theories, one begins by rewriting the equations of motion as a
zero curvature condition, i.e., as the compatibility condition for a linear
system with spectral parameter (Lax pair). In the Hamiltonian context one
then studies the Poisson brackets between $L$-matrices (the spatial part of
the Lax pair). In the most common cases, this leads to a Lie-Poisson structure
with an numerical (field independent) antisymmetric $r$-matrix
obeying the classical Yang-Baxter equation.
Theories for which the Poisson brackets between $L$-matrices are of this
form are commonly called ultralocal \cite{FaTa} because the rhs.
contains only the delta function $\delta(x-y)$ but not
its derivatives.
An important generalization of this structure to certain
non-ultralocal models, namely those for which the rhs.
contains, apart from the delta function, its first derivative (i.e., a
classical Schwinger term), has been developed by Maillet \cite{Mai1}.
This generalization leads to $r$-matrices which are no longer antisymmetric and
may also depend on the dynamical variables, i.e. they will obey a dynamical
Yang-Baxter equation.
\begin{footnote}{
Some of the most important
non-ultralocal models, such as the $O(N)$-sigma model \cite{DEM,Mai2},
the principal chiral model \cite{DEM,Mai3} and the complex sine-Gordon model
\cite{Mai1}, are known to fit into this scheme.}
\end{footnote}

\noindent We begin by briefly reviewing the results on the current algebra
structure of integrable non-linear sigma models, (which are precisely those
defined on Riemannian symmetric spaces $M$), derived in a previous paper
\cite{FLS}\cite{L}.

\noindent Technically, we require that $M$ is the
quotient space of some connected Lie group $G$, with Lie algebra ${\gotig}$,
modulo some compact subgroup $H\subset G$, with Lie algebra
${\gotigh}\subset{\gotig}$.
\begin{footnote}{
Furthermore, there exists an $Ad(H)$ invariant subspace $\goth m$  of
${\goth g}$ with:
$ [\mbox{\goth h}\, , \mbox{\goth h}\,] \smsubset \mbox{\goth h}~,~~
 [\mbox{\goth h}\, , \mbox{\goth m}\,] \smsubset \mbox{\goth m}~,~~
 [\mbox{\goth m}\, , \mbox{\goth m}\,] \smsubset \mbox{\goth h}~,
$\newline
such that $\mbox{\goth g}\,$ is the (vector space) direct sum of
$\mbox{\goth h}\,$ and $\mbox{\goth m}\,$:~
$ \mbox{\goth g}\, = \mbox{\goth h}\, \oplus \mbox{\goth m}~~~. \label{eq:DD}$
\newline
Moreover we suppose that this decomposition is orthogonal.}
\end{footnote}
Then $G$ acts on $M$ by isometries and the $G$
invariance of the action leads to a conserved Noether current
$j_\mu$, with values in $\gotig^\ast$ (the dual of $\gotig\,$). The Poisson
algebra of these currents can be written in a closed form after
introducing a scalar field $j$ with values in the second symmetric tensor
power $S^2(\gotig^*)$ of $\gotig^\ast$. To do so it is convenient to introduce
a basis
$(T_a)$ of ${\gotig}$, with structure constants $f_{ab}^c$, together with the
corresponding dual basis $(T^a)$ of $\gotig^\ast$, and to expand $j_\mu$
and $j$ into components:
$$
j_\mu = j_{\mu,a}T^a \;\;,\;\;\; j = j_{ab}T^a \vee T^b.
$$
With this notation the current algebra takes the form
\cite{FLS}
\begin{eqnarray}
 \{ j_{0,a}(x) , j_{0,b}(y) \}
 &=& - \, f_{ab}^c , j_{0,c}(x) \, \delta(x-y)~~,      \label{eq:CA1} \\
 \{ j_{0,a}(x) , j_{1,b}(y) \}
 &=& - \, f_{ab}^c \, j_{1,c}(x) \, \delta(x-y) \,
     + \, j_{ab}(y) \, \delta^\prime (x-y)~~,    \label{eq:CA2} \\
 \{ j_{0,a}(x) , j_{bc}(y) \}
 &=& - \left( f_{ab}^d \, j_{cd}(x) + f_{ac}^d \, j_{bd}(x) \right)
                                    \delta(x-y)~~,      \label{eq:CA3}
\end{eqnarray}
All other Poisson brackets being zero.

\noindent The current algebra (\ref{eq:CA1})-(\ref{eq:CA3}) governs the
canonical structure of these models and therefore is the key feature to
analyze the integrable structure.
The integrability of a classical two-dimensional field theory is expressed
through the possibility to rewrite its equations of motion as a zero curvature
condition~\cite{FaTa}.
\begin{equation}
 \partial_\mu L_\nu - \partial_\nu L_\mu + [L_\mu,L_\nu]~=~0~~~,
                                                                 \label{eq:ZC2}
\end{equation}
where $L_\mu$ is a function on two-dimensional space-time taking
values in the Lie algebra
$\mbox{\goth g}\,$, and depending on an additional spectral
parameter $\lambda$. In the present case, $L_\mu$ is a
$\lambda$-dependent linear combination of the current $j_\mu$ and its (Hodge)
dual $\, \epsilon_{\mu\nu} j^\nu$, namely
\begin{equation}
 L_\mu~=~{2 \over 1\!-\!\lambda^2} \left( j_\mu + \lambda
                      \epsilon_{\mu\nu} j^\nu \right)~~~.        \label{eq:LM1}
\end{equation}
The $L$-matrix $L$ is defined to be the spatial component
of the flat vector potential $L_\mu$, expanded into components
$ L~=~L_a \, T^a~~~,$
In these terms, the Poisson bracket algebra of the $L$-matrix
follows directly from
combining the current algebra (\ref{eq:CA1})--(\ref{eq:CA3}) with the
definition (\ref{eq:LM1}) of $L$\cite{FBLS}:
\begin{eqnarray}
 \{ L_a(x,\lambda) \, , \, L_b(y,\mu) \}
 &\!=\!& {2 f_{ab}^c \over \lambda\!-\!\mu} \,
         \Bigl( {\mu^2 \over 1\!-\!\mu^2} \, L_c(x,\lambda) \, - \,
                {\lambda^2 \over 1\!-\!\lambda^2} \, L_c(x,\mu) \Bigr) \;
         \delta(x-y)                                        \nonumber \\
 &\!   & + \; {4 \over (1\!-\!\lambda^2)(1\!-\!\mu^2)} \,
         \Bigl( \mu j_{ab}(x) + \lambda j_{ab}(y) \Bigr) \;
         \delta^\prime(x-y)~,~~~~~~                   \label{eq:PB1a} \\
 \{ L_a(x,\lambda) \, , \, j_{bc}(y) \}
 &\!=\!& - \, {2\lambda \over 1\!-\!\lambda^2} \,
         \Bigl( f_{ab}^d \, j_{cd}(x) + f_{ac}^d \, j_{bd}(x) \Bigr) \;
         \delta(x-y)~.~~~~~~                          \label{eq:PB2a} \\
 \{ j_{ab}(x) \, , \, j_{cd}(y) \} &\!=\!& 0~.~~~~~~            \label{eq:PB3a}
\end{eqnarray}

\noindent Our next goal now is to exhibit
the full Lie-Poisson structure of the theory -- which in terms of components
is contained in eqns (\ref{eq:PB1a}) and (\ref{eq:PB2a})
 -- and to discuss the algebraic constraints resulting from
the Jacobi identity for Poisson brackets. This requires, first of all, a
slight modification of the tensor notation of \cite{FaTa}
to deal with arbitrary tensor powers $U(\gotg\,)^{\otimes n}$.
\begin{footnote}{
We use the following notation for embeddings:
$ U(\gotg\,) \otimes U(\gotg\,) \longrightarrow U(\gotg\,)^{\otimes n}~,~~
u \smotimes v \longmapsto (u \smotimes v)_{kl}
{}~=~1 \smotimes \!\ldots\! \smotimes u \smotimes \!\ldots\! \smotimes
                                    v \smotimes \!\ldots\! \smotimes 1
$
(with $u$ appearing in the $k^{\rm th}$ place and $v$ appearing in the
$l^{\rm th}$ place, $1 \leq k,l \leq n$), respectively.}
\end{footnote}

\noindent In this notation, the Poisson brackets for the $L$'s and $j$'s given
by eqns (\ref{eq:PB1a})--(\ref{eq:PB3a}) read
\begin{eqnarray}
 \{ L_k(x_k,\lambda_k) \, , \, L_l(x_l,\lambda_l) \}
 &\!=\!& \left[ \, {2 \, C_{kl} \over \lambda_k\!-\!\lambda_l} \; , \;
         {\lambda_l^2 \over 1\!-\!\lambda_l^2} \, L_k(x_k,\lambda_k) \, + \,
         {\lambda_k^2 \over 1\!-\!\lambda_k^2} \, L_l(x_l,\lambda_l) \,
         \right] \, \delta(x_k-x_l)                         \nonumber \\[0.1cm]
 &\!   & + \; {4 \over (1\!-\!\lambda_k^2)(1\!-\!\lambda_l^2)} \,
         \Bigl( \lambda_l j_{kl}(x_k) + \lambda_k j_{kl}(x_l)
         \Bigr) \, \delta^\prime(x_k-x_l)~,~~~~~~     \label{eq:PB1f} \\[0.2cm]
 \{ L_k(x_k,\lambda_k) \, , \, j_{lm}(x_l) \}
 &\!=\!& {2\lambda_k \over 1\!-\!\lambda_k^2} \,
         \Bigl[ \, C_{kl} + C_{km} \, , \, j_{lm}(x_k) \, \Bigr] \;
         \delta(x_k-x_l)~,~~~~~~                      \label{eq:PB2e} \\[0.2cm]
 \{ j_{kl}(x_k) \, , \, j_{mn}(x_m) \} &\!=\!& 0~.~~~~~~        \label{eq:PB3e}
\end{eqnarray}
with the Casimir tensor $C_{kl} = \eta^{ab} (T_a)_k (T_b)_l$
If we now pass
from the $L$-matrix $L(x,\lambda)$ to the corresponding Lax operator
$D(x,\lambda)$, defined as\newline
$D(x,\lambda)~=~{\partial \over \partial x} \, + \, L(x,\lambda)~~~.
          \label{eq:LO} $ \newline
the Poisson brackets of the $D$'s are, by definition, the same
as the ones for the $L$'s, but $D$ being a differential operator, the
inhomogeneous classical Schwinger terms on the rhs.\ of eqn (\ref{eq:PB1f})
can be absorbed into commutators (the commutator of a $D$
with a delta function produces, among other things, the derivative of a delta
function).
Therefore, the Poisson brackets
for the $D$'s,
take the following, much more transparent form:
\setlength{\arraycolsep}{0pt}\vfill\eject
\begin{eqnarray}
 & &\{ D_k(x_k,\lambda_k) \, , \, D_l(x_l,\lambda_l) \}     \nonumber \\
 & &\hspace{2.1cm} =~\;\left[ \, d_{kl}(x_k,\lambda_k\,;x_l,\lambda_l) \, , \,
                                 D_k(x_k,\lambda_k) \, \right] \, - \,
                       \left[ \, d_{lk}(x_l,\lambda_l\,;x_k,\lambda_k) \, , \,
                                 D_l(x_l,\lambda_l) \,
                       \right]~,~~~~~~                \label{eq:PB1g} \\
 & &\{ D_k(x_k,\lambda_k) \, , \, d_{lm}(x_l,\lambda_l\,;x_m,\lambda_m) \}
                                                            \nonumber \\
 & &\hspace{2.1cm} =~\;\left[ \, c_{kl}(x_k,\lambda_k\,;x_l,\lambda_l) +
                                 c_{km}(x_k,\lambda_k\,;x_m,\lambda_m) \, , \,
                                 d_{lm}(x_l,\lambda_l\,;x_m,\lambda_m) \,
                       \right]~,~~~~~~                \label{eq:PB2g} \\
 & &\{ d_{kl}(x_k,\lambda_k\,;x_l,\lambda_l) \, , \,
       d_{mn}(x_m,\lambda_m\,;x_n,\lambda_n) \}~~=~~0~,         \label{eq:PB3g}
\end{eqnarray}
with
\begin{equation}
 d(z,\lambda,\mu)
 =~{2\lambda\mu \over 1\!-\!\lambda\mu}
   \left( {C \over \lambda\!-\!\mu} \, - \,
          {2 \, j(z) \over \lambda (1\!-\!\mu^2)} \right)~~~,     \label{eq:d1}
\end{equation}
and
\begin{equation}
 c(z,\lambda,\mu)~=~{2\lambda \over 1\!-\!\lambda^2} \, C~~~.     \label{eq:c1}
\end{equation}
These equations show that the $D$'s and $d$'s
generate an algebra which closes under Poisson brackets, because $c$ is a
numerical (i.e., field independent) matrix (cf.\ (\ref{eq:c1})). In the theory
of non-ultralocal integrable models of the type considered here, this algebra
plays a central role: it is the analogue of the classical Yang-Baxter algebra
which is relevant to ultralocal integrable models.
The structure of the
abstract algebra is (at least partially) reflected in the $c$-matrix,
while the $D$'s and $d$'s define a concrete representation of that
algebra by functionals on the phase space of the theory, according to
eqns (\ref{eq:PB1g})--(\ref{eq:PB3g}).
The investigation of the mathematical structures that underly this new algebra
is still a completely open subject. The first step would be to identify its
defining relations, i.e., the analogue of the Jacobi identity for Lie algebras
or the classical Yang-Baxter equation for classical Yang-Baxter algebras: they
should include a structure equation for the $c$-matrix which we suspect, once
again, to be quadratic.
Here, we just want to analyze the consequences of the fact
that eqns (\ref{eq:PB1g})--(\ref{eq:PB3g})
must be consistent with the Jacobi identity for Poisson brackets.

Moreover, we shall find it convenient to use the following general notation,
for $\;t \smin {U(\gotg)} \otimes {U(\gotg)}\;$ and $\, k,l,m \,$ all distinct,
\begin{equation}
 {\rm YB}(t)_{klm}~~=~~[ \, t_{kl} \, , \, t_{km} \, ] \; + \;
                       [ \, t_{kl} \, , \, t_{lm} \, ] \; - \;
                       [ \, t_{km} \, , \, t_{ml} \, ]~~~.       \label{eq:DYB}
\end{equation}
Turning to the verification of the Jacobi identity, we find from eqn
(\ref{eq:PB1g}), the following relation
\cite{BaVi,Mai1}:
\begin{eqnarray}
\lefteqn{\{ \, D_k \, , \, \{ \, D_l \, , \, D_m \, \} \, \}~+~{\rm cyclic}}
                                                    \hspace{0.5cm} \nonumber \\
 &=& [ \; D_k \; , \; {\rm YB}(d)_{klm} \, + \,
                      [ \, c_{lk} + c_{lm} \, , \, d_{km} \, ] \, - \,
                      [ \, c_{mk} + c_{ml} \, , \, d_{kl} \, ] \; ]
     ~+~{\rm cyclic}~.~~~~~~                                    \label{eq:DDD2}
\end{eqnarray}
Thus the rhs.\ of this expression must vanish in order for the Jacobi identity
to be satisfied. But actually, more than this is true: namely, we have
\begin{equation}
 {\rm YB}(d)_{klm} \, + \, [ \, c_{lk} + c_{lm} \, , \, d_{km} \, ] \, - \,
                           [ \, c_{mk} + c_{ml} \, , \, d_{kl} \, ]~~=~~0~~~,
                                                                \label{eq:YBE4}
\end{equation}
as can be checked by an explicit calculation
this confirms, for the class of models
under consideration here, the validity of the ``extended dynamical
Yang-Baxter relation'' postulated by Maillet \cite{Mai1}.
Similarly, we obtain from eqns (\ref{eq:PB2g})--(\ref{eq:PB3g})
\begin{eqnarray}
\lefteqn{\{ \, D_k \, , \, \{ \, D_l \, , \, d_{mn} \, \} \, \}~+~
         \{ \, D_l \, , \, \{ \, d_{mn} \, , \, D_k \, \} \, \}~+~
         \{ \, d_{mn} \, , \, \{ \, D_k \, , \, D_l \, \} \, \}}
                                                    \hspace{0.5cm} \nonumber \\
 &=& [ \; d_{mn} \; , \;
     [ \, c_{km} + c_{kn} \, , \, c_{lm} + c_{ln} \, ]~-~
     [ \, c_{km} + c_{kn} \, , \, d_{kl} \, ]~+~
     [ \, c_{lm} + c_{ln} \, , \, d_{lk} \, ] \; ]~.~~~~~~~~     \label{eq:DDd}
\end{eqnarray}
Again the rhs.\ of this expression must vanish in order for the Jacobi identity
to be satisfied, and again this can be checked by an explicit calculation
Finally, in the remaining combinations
\[
  \{ \, D_i \, , \, \{ \, d_{kl} \, , \, d_{mn} \, \} \, \}~+~
  \{ \, d_{kl} \, , \, \{ \, d_{mn} \, , \, D_i \, \} \, \}~+~
  \{ \, d_{mn} \, , \, \{ \, D_i \, , \, d_{kl} \, \} \, \}
\vspace{-0.2cm}
\]
and
\[
  \{ \, d_{ij} \, , \, \{ \, d_{kl} \, , \, d_{mn} \, \} \, \}~+~
  \{ \, d_{kl} \, , \, \{ \, d_{mn} \, , \, d_{ij} \, \} \, \}~+~
  \{ \, d_{mn} \, , \, \{ \, d_{ij} \, , \, d_{kl} \, \} \, \}
\vspace{0.2cm}
\]
each term vanishes separately.

\end{document}